\pdfoutput=1
\documentclass[11pt]{article}

\usepackage[]{packages/acl}

\usepackage{times}
\usepackage{latexsym}
\usepackage[T1]{fontenc} 
\usepackage[utf8]{inputenc}
\usepackage{microtype} 

\usepackage[show]{packages/chato-notes} 
\usepackage[pdftex]{graphicx}           
\usepackage{subcaption}                 
\usepackage{booktabs}                   
\usepackage{amsmath}                    
\usepackage{amsfonts}                   
\usepackage{nicefrac}                   
\usepackage{multirow}                   
\usepackage{pifont}                     
\usepackage{enumitem}                   
\usepackage{cleveref}                   

\setlist[itemize,enumerate]{itemsep=1pt, topsep=1pt}
\crefformat{footnote}{#2\footnotemark[#1]#3}
\newcommand{\at}{\texttt{@}}%
\newcommand{\cls}{\textsc{[cls]}}%
\newcommand{\house}{\includegraphics[scale=0.06]{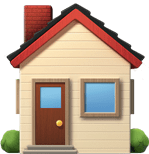}}
\newcommand{\money}{\includegraphics[scale=0.06]{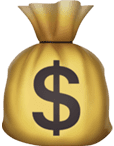}}
\newcommand{\printer}{\includegraphics[scale=0.06]{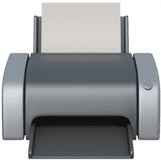}}
\newcommand{\family}{\includegraphics[scale=0.06]{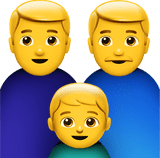}}
\newcommand{\scale}{\includegraphics[scale=0.06]{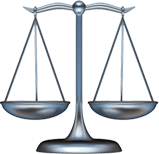}}
\newcommand{\plane}{\includegraphics[scale=0.06]{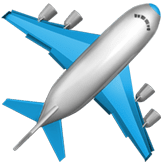}}
\newcommand{\old}{\includegraphics[scale=0.06]{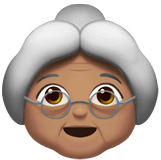}}
\colorlet{shade}{gray!90}

\title{Finding the Law: Enhancing Statutory Article Retrieval via Graph \\Neural Networks}

\author{
Antoine Louis \hspace{0.8cm} Gijs van Dijck \hspace{0.8cm} Gerasimos Spanakis \\
Law \& Tech Lab, Maastricht University \\
{\small \texttt{\{a.louis, gijs.vandijck, jerry.spanakis\}@maastrichtuniversity.nl}} \\
}

\begin{document}
\maketitle

\begin{abstract}
Statutory article retrieval (SAR), the task of retrieving statute law articles relevant to a legal question, is a promising application of legal text processing. In particular, high-quality SAR systems can improve the work efficiency of legal professionals and provide basic legal assistance to citizens in need at no cost. Unlike traditional ad-hoc information retrieval, where each document is considered a complete source of information, SAR deals with texts whose full sense depends on complementary information from the topological organization of statute law. While existing works ignore these domain-specific dependencies, we propose a novel graph-augmented dense statute retriever (G-DSR) model that incorporates the structure of legislation via a graph neural network to improve dense retrieval performance. Experimental results show that our approach outperforms strong retrieval baselines on a real-world expert-annotated SAR dataset.\footnote{Our source code is available at \url{https://github.com/maastrichtlawtech/gdsr}.}
\end{abstract}

\section{Introduction \label{sec:introduction}}
Today, the high cost of legal expertise prevents less fortunate people from understanding and reacting to legal issues that may arise \citep{long2019global}. In recent years, an increasing number of works have focused on legal text processing \citep{zhong2020how} with the intent to assist legal practitioners and citizens while reducing legal costs and improving equal access to justice for all. Statutory article retrieval (SAR), the task of retrieving statute law articles relevant to a legal question, marks the first and one of the most crucial steps in any legal aid process. Our goal is to help reduce the gap between people and the law by improving SAR systems that could provide citizens with the first component of a free professional legal aid service.

\begin{figure}[t]
    \centering
    \includegraphics[width=0.86\linewidth]{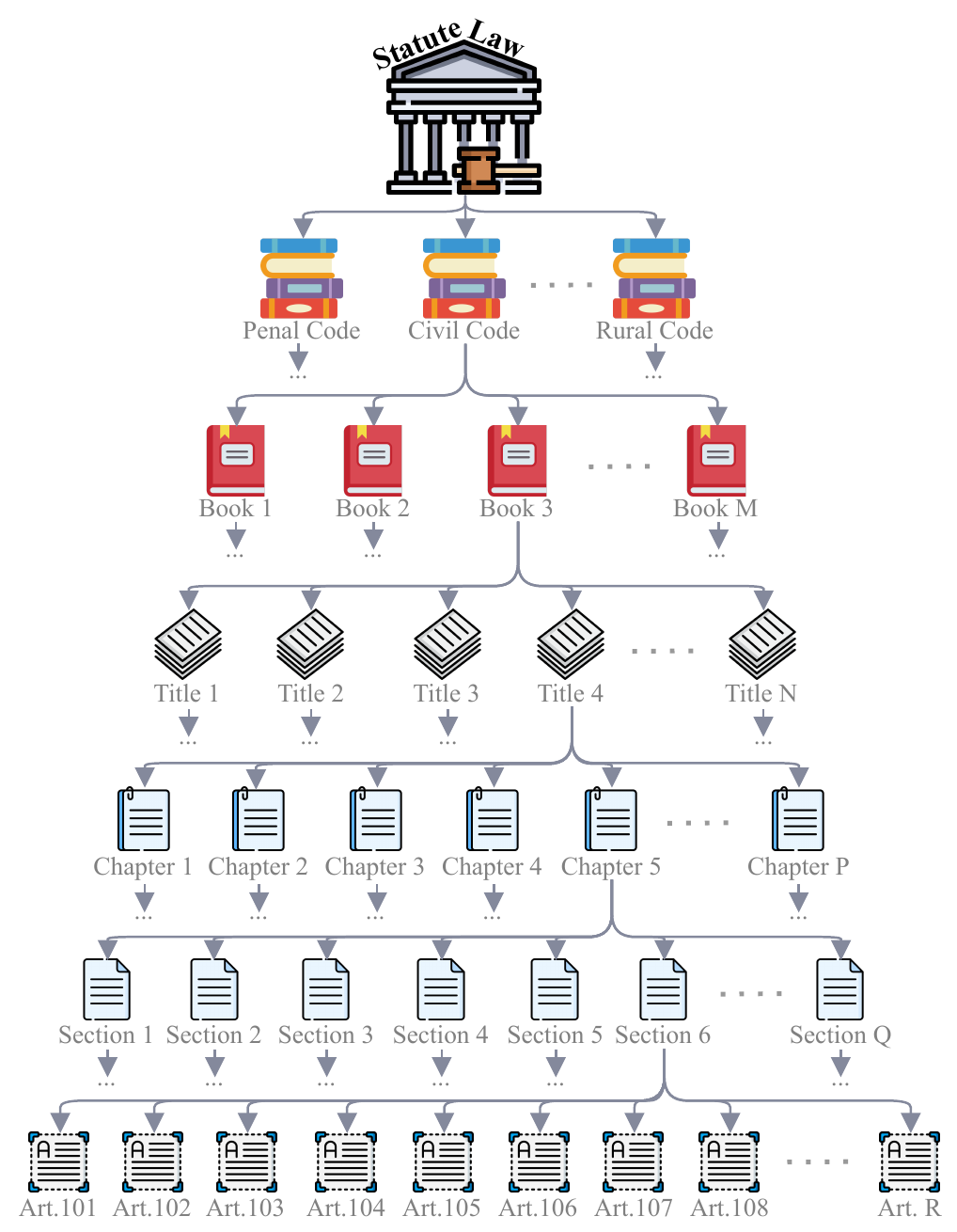}
    \caption{Illustration of the hierarchical organization of statute law. Each law code is structured into books, titles, chapters, and sections. The deeper the divisions, the closer the legal concepts of the articles below them.}
    \label{fig:hierarchy}
\end{figure}

Prior work has addressed SAR with standard information retrieval approaches such as term-based models or dense embedding-based models \citep{kim2019statute,nguyen2021jnlp}. While good performance has been achieved, these approaches rely on the flawed assumption that articles are complete and independent sources of information. In reality, statute law is an ensemble of \textsl{interdependent} rules meticulously organized into different codes, books, titles, chapters, and sections, as illustrated in \Cref{fig:hierarchy}. Each level in the structure of legislation comes with a unique heading that informs about the content of the articles below it. An article takes on its whole meaning only when considered at its rightful place in the structure with the complementary information from its neighboring articles.

This work shows that such a structure can be highly beneficial for retrieving statutes. We propose a graph-augmented dense statute retriever (G-DSR) model that leverages the topological structure of legislation to enhance the article content information. Specifically, the proposed model extends the document encoder of a dense retriever with a graph neural network to learn knowledge-rich cross-article representations. Similar to previous work, we adopt a contrastive learning strategy to optimize the similarity between the representations of relevant query-article pairs.

The contributions of this paper are threefold:
\begin{itemize}
    \item We propose a graph-augmented dense retriever model for statutory article retrieval that explicitly utilizes the topological organization of statute law to enrich the article information.
    \item We conduct empirical evaluations on our model and demonstrate improvements over strong retrieval baselines.
    \item We perform ablation studies on various model components and training strategies to understand the impact of several design options on the effectiveness of our model.
\end{itemize}

\section{Preliminaries \label{sec:preliminaries}}
In this section, we formally introduce the task of \textsl{statutory article retrieval} and discuss the specific difficulties associated with it. We then explain how we identify the structure of legislation as an essential consideration in SAR.

\paragraph{Problem formulation.}
Given a simple legal question, such as "\textsl{Who should pay for the construction of the common wall?}", SAR aims to return one or several relevant articles from the legislation. Formally speaking, a SAR system can be expressed as a function $R: (q, \mathcal{C}) \mapsto \mathcal{F}$ that takes as input a question $q$ along with a corpus of articles $\mathcal{C}=\{a_1, a_2, \cdots, a_N\}$, and returns a much smaller filter set $\mathcal{F} \subset \mathcal{C}$ of the supposedly relevant articles, ranked by decreasing order of relevance. For a fixed $k=\left|\mathcal{F}\right|\ll|\mathcal{C}|$, the retriever can be evaluated in isolation with multiple rank-based metrics. Most modern retriever systems follow a two-stage retrieval approach \citep{guo2016deep,hui2017pacrr,mcdonald2018deep}, where a \textsl{pre-fetcher} first aims to return all relevant documents in the filter set $\mathcal{F}$ and a \textsl{re-ranker} then attempts to make more relevant documents in $\mathcal{F}$ appear before less relevant ones. In this work, we focus our research on improving the pre-fetcher component for SAR.

\paragraph{Challenges.}
SAR comes with two core challenges that make the task unique compared to traditional information retrieval. First, the statutes to be retrieved are written in a language that dramatically differs from the ordinary plain language used in the questions. The legal language uses a specialized jargon known for its frequent and deliberate use of formal words, Latin phrases, lengthy sentences, and expressions with flexible meanings \citep{charrow1990legal}. Second, statutory articles are long text sequences that may reach several thousand words. This implies overcoming the maximum input length limit of 512 tokens imposed by BERT-based models, which have recently become the standard in neural information retrieval due to their effectiveness.

\paragraph{Structure of legislation.}
The legislation comes with a well-thought-out organization of its written rules to facilitate access to provisions covering a given subject \citep{onoge2015structure}. This organization is established in a hierarchical manner, where higher-level divisions cover broad legal domains while lower-level divisions deal with specific legal concepts. To examine the importance of this hierarchy in the SAR process, we conduct a preliminary investigation in which we study the reasoning legal experts follow when performing the task. We summarize these experts' approach in \Cref{app:experts_approach}. We observe that legal experts rely heavily on the structure of law when retrieving relevant articles to a legal question, which indicates that the different divisions' headings in the legislation carry valuable information that retrieval systems should consider. Additionally, we analyze the degree to which neighboring articles cover related subjects in \Cref{app:neighboring_articles} and find high levels of similarities, which suggests that information from neighboring articles should be considered to capture an article's whole meaning.

\begin{figure*}[t]
    \centering
    \includegraphics[width=1.0\linewidth]{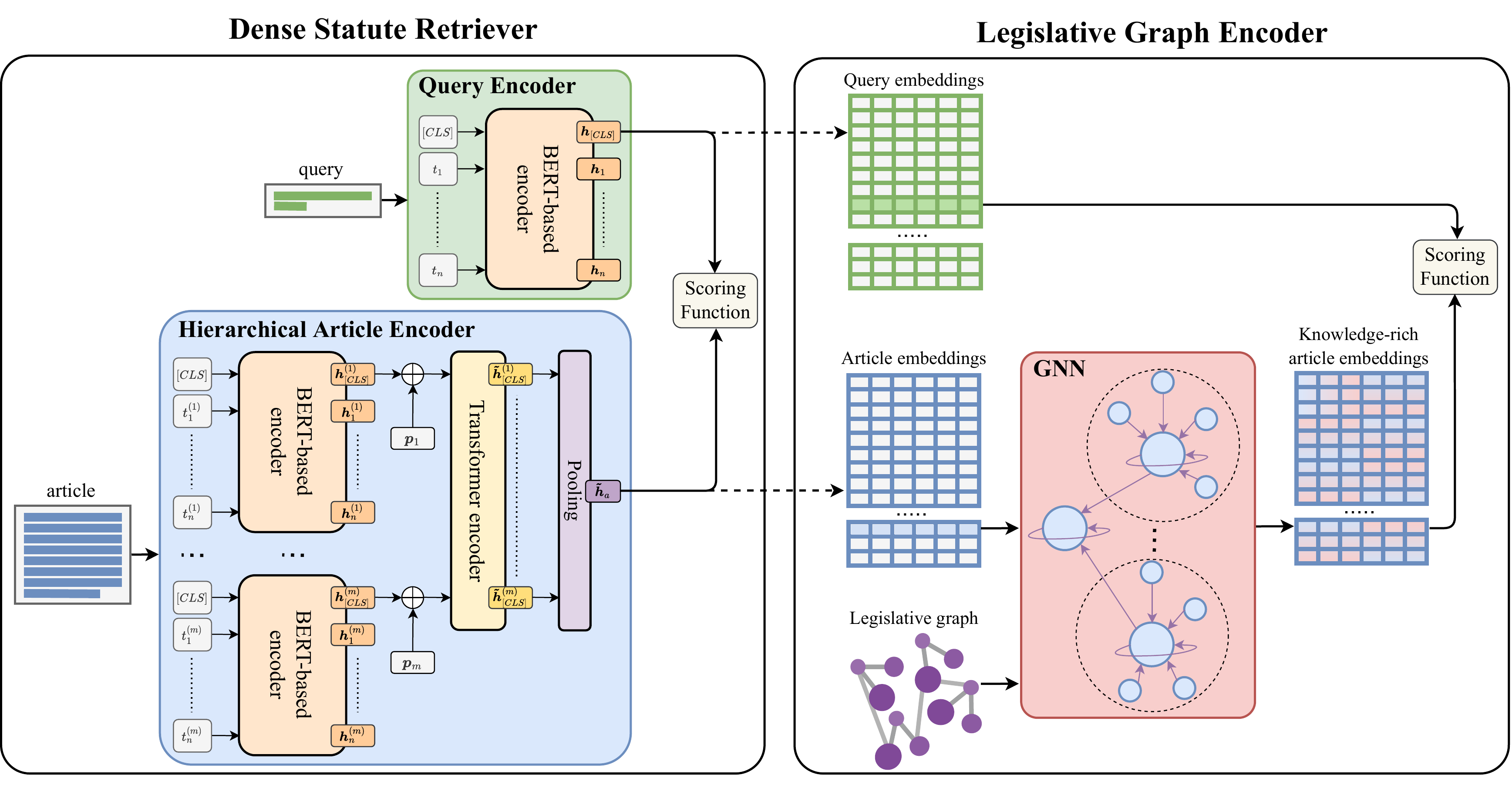}
    \caption{\textbf{An illustration of the graph-augmented dense statute retriever (G-DSR) model}. G-DSR consists of two main building blocks that are trained independently. \textbf{Left:} The dense statute retriever (DSR) first learns high-quality low-dimensional embedding spaces for both the queries and articles such that relevant query-article pairs appear closer than irrelevant ones in those vector spaces. \textbf{Right:} The legislative graph encoder (LGE) then learns to enrich the article representations by aggregating information from the organization of statute law.}
    \label{fig:model}
\end{figure*}

\section{Approach \label{sec:approach}}
In this section, we present a new general approach for SAR that learns to retrieve relevant statutes by using both the textual semantic information from articles and the structural graph information from the legislation. Our model, called graph-augmented dense statute retriever (G-DSR), consists of two main building blocks, as depicted in \Cref{fig:model}, that are trained independently with the same objective. We first describe the dense retriever component of our approach in \Cref{subsec:dsr} and then explain how our legislative graph encoder builds upon it in \Cref{subsec:lge}.

\subsection{Dense Statute Retriever \label{subsec:dsr}}
Our approach's first component, called dense statute retriever (DSR), aims to learn high-quality low-dimensional embedding spaces for questions and articles so that relevant question-article pairs appear closer than irrelevant ones in those spaces. Below, we review the overall architecture of the retriever and detail the design of its query and article encoders. We then describe the contrastive learning strategy we employ and choice of negative pairs.

\paragraph{Bi-encoder.}
We use the widely adopted bi-encoder architecture \citep{bromley93signature} as the foundation of our dense retriever. The latter maps queries and articles into dense vector representations and calculates a relevance score $s: (q,a) \mapsto \mathbb{R}_+$ between query $q$ and article $a$ by the similarity of their embeddings, i.e.,
\begin{equation}
    s(q,a) = \mathrm{sim}\left(E_Q^{\boldsymbol{\theta}}(q), E_A^{\boldsymbol{\phi}}(a)\right),
\end{equation}
where $E_Q^{\boldsymbol{\theta}}(q), E_A^{\boldsymbol{\phi}}(a) \in \mathbb{R}^{d}$ denote the query and article embeddings respectively, and $\mathrm{sim}: \mathbb{R}^d \times \mathbb{R}^d \mapsto \mathbb{R}$ is a similarity function such as cosine or dot-product.

\paragraph{Query encoder.}
To encode the queries, we feed them into a BERT-based \citep{devlin2019bert} model $E_Q^{\boldsymbol{\theta}}: \mathcal{W}^{n} \mapsto \mathbb{R}^{d}$ with weights $\boldsymbol{\theta}$, that maps an input sequence of $n$ tokens from vocabulary $\mathcal{W}$ to $d$-dimensional real-valued token embeddings. We take the last layer's \cls{} token representation as the query embedding, i.e.,
\begin{equation}
    E_Q^{\boldsymbol{\theta}}(q) = \text{BERT}_{\cls}(q).
\end{equation}

\paragraph{Hierarchical article encoder.}
Since statutory articles may be longer than the maximum input length of a standard BERT-based encoder, we use a hierarchical variation that can process longer textual sequences \citep{pappagari2019hierarchical,zhang2019hibert,yang2020beyond}. Each article $a$ is first split into smaller text passages $[p_1, p_2,\cdots,p_m]$, where a passage $p_i$ is a sequence of tokens $[t_{1}^{(i)},t_{2}^{(i)},\cdots, t_{n}^{(i)}]$ with $n \leq 512$. These passages are then independently passed through a shared BERT-based model to extract a list of \textsl{context-unaware} passage representations using the respective \cls{} token embeddings, as illustrated in \Cref{fig:model}. Next, the hierarchical model sums the \cls{} token representations of each passage with learnable passage position embeddings and feeds the resulting representations into a small Transformer encoder to make them aware of the surrounding passages. The final article representation is computed through a pooling operation over the \textsl{context-aware} passage representations, i.e.,
\begin{equation}
    E_A^{\boldsymbol{\phi}}(a) = \mathrm{pool}\left(\left[\boldsymbol{\tilde{h}}_{\cls}^{(1)}, \cdots,\boldsymbol{\tilde{h}}_{\cls}^{(m)}\right]\right),
\end{equation}
where $\boldsymbol{\tilde{h}}_{\cls}^{(i)} \in \mathbb{R}^{d}$ is the contextualized embedding of passage $p_i$, and  $\mathrm{pool}: \mathbb{R}^{m \times d} \mapsto \mathbb{R}^{d}$ is either mean or max pooling.

\paragraph{Contrastive learning.}
The training objective of the bi-encoder is to learn effective embedding functions $E_Q^{\boldsymbol{\theta}}(\cdot)$ and $E_A^{\boldsymbol{\phi}}(\cdot)$ such that relevant question-article pairs have a higher similarity than irrelevant ones. Let $\mathcal{D}=\{\langle q_{i}, a_{i}^{+}\rangle\}_{i=1}^{N}$ be the training data where each of the $N$ instances consists of a query $q_{i}$ associated with a relevant article $a_{i}^{+}$. By sampling a set of negative articles $\mathcal{A}_{i}^{-}$ for each question $q_i$, we can create a training set $\mathcal{T}=\{\langle q_{i}, a_{i}^{+}, \mathcal{A}_{i}^{-}\rangle\}_{i=1}^{N}$. For each training instance in $\mathcal{T}$, we contrastively optimize the negative log-likelihood of the positive article against the negative ones, i.e.,
\begin{equation}
\label{eq:loss}
    L\left(q_{i}, a_{i}^{+}, \mathcal{A}_{i}^{-}\right) = -\log \frac{e^{s(q_i,a_{i}^{+})/\tau}}{\sum_{a \in \mathcal{A}_{i}^{-} \cup \{a_{i}^{+}\}} e^{s(q_{i}, a)/\tau}},
\end{equation}
where $\tau > 0$ is a temperature parameter to be set.

\paragraph{Negatives.}
We consider two types of negative examples: (i) in-batch \citep{chen2017sampling, henderson2017efficient}, i.e., articles paired with the other questions from the same mini-batch, and (ii) BM25, i.e., top articles returned by BM25 that are not relevant to the question.

\subsection{Legislative Graph Encoder \label{subsec:lge}}
Our approach's second component, called Legislative Graph Encoder (LGE), aims to enrich article representations given by the trained retriever's article encoder by fusing information from a legislative graph. Below, we elaborate on the legislative graph construction and the graph training process.

\paragraph{Graph construction.}
To leverage the hierarchical organization of statute law, we formalize the latter as a tree structure consisting of two types of node: (i) \textsl{section} nodes, which are titled structural units that represent the consecutive divisions in codes of law (i.e., the headings of the books, titles, chapters, and sections), and (ii) \textsl{article} nodes, which are textual content units that represent the different statutory articles. As illustrated in \Cref{fig:hierarchy}, the edges represent the hierarchical connections between section and article nodes. Formally, such a tree can be represented as a directed acyclic graph $\mathcal{G} = (\mathcal{V}, \mathcal{E})$, with $\mathcal{V}$ as the node set and $\mathcal{E} \subseteq \mathcal{V} \times \mathcal{V}$ as the edge set.

\paragraph{Node feature initialization.}
Nodes in $\mathcal{V}$ are commonly associated with $d$-dimensional features. We apply the article encoder $E_A^{\boldsymbol{\phi}}(\cdot)$ from the trained bi-encoder to encode the semantic information of nodes (i.e., section headings and article contents) offline and use the resulting embeddings as the initial node features $\mathbf{X} \in \mathbb{R}^{|\mathcal{V}| \times d}$.

\paragraph{Node feature update.}
To fuse the information of node features using the graph structure, we use a graph neural network (GNN). Such a model consists of a stack of neural network layers, where each layer aggregates local neighborhood information (i.e., features of neighbors) around each node and then passes this aggregated information on to the next layer. Generally speaking, a GNN takes as inputs the feature matrix $\mathbf{X}$ and the graph's adjacency matrix $\mathbf{A} \in \mathbb{R}_{+}^{|\mathcal{V}| \times |\mathcal{V}|}$, with $\mathbf{A}_{i,j}$ as the edge weight between nodes $i$ and $j$, and produces a node-level output $\mathbf{Z} \in \mathbb{R}^{|\mathcal{V}| \times d}$ that captures each node's structural properties. Every GNN layer can be written as a non-linear function
\begin{equation}
    \mathbf{H}^{(l+1)} = f(\mathbf{H}^{(l)}, \mathbf{A}),
\end{equation}
with $\mathbf{H}^{0} = \mathbf{X}$ and $\mathbf{H}^{L} = \mathbf{Z}$, $L$ being the number of layers. In its simplest form, the layer-wise propagation rule is such that
\begin{equation}
\label{eq:gnn_simple}
    f(\mathbf{H}^{(l)}, \mathbf{A}) = \sigma(\mathbf{A}\mathbf{H}^{(l)}\mathbf{W}^{(l)}),
\end{equation}
where $\mathbf{W}^{(l)}$ is the input linear transformation's weight matrix for the $l$-th neural network layer and $\sigma(\cdot)$ is a non-linear activation function. We propose to use a 3-layer GATv2 network \citep{brody2022attentive}, a variant of GAT \citep{velickovic2018graph} that has the ability to learn the strength of connection between neighboring nodes through a \textsl{dynamic} attention mechanism. Formally, a GATv2 layer updates a node's hidden state as follows
\begin{equation}
    \boldsymbol{h}_{i}^{(l+1)}=\sigma\left(\sum_{j \in \mathcal{N}(i)} \alpha_{i j}^{(l)} \mathbf{W}^{(l)} \boldsymbol{h}_{j}^{(l)}\right),
\end{equation}
where $\mathcal{N}(i)$ is the set of first-order neighbors of node $i$, and $\alpha_{i j}^{(l)}$ are normalized attention coefficients indicating the importance of node $j$'s features to $i$ in the $l$-th layer. The latter are computed based on the features of the connected nodes using an attention function $\mathrm{att}: \mathbb{R}^d \times \mathbb{R}^d \mapsto \mathbb{R}$ such that
\begin{equation}
    \alpha_{i j}^{(l)} = \mathrm{softmax}\left(\sigma(\mathrm{att}(\boldsymbol{h}_{i}^{(l)}, \boldsymbol{h}_{j}^{(l)})\right).
\end{equation}

\paragraph{Learning process.}
To optimize the GNN parameters, we adopt the same contrastive learning strategy used to train the bi-encoder. Since graph $\mathcal{G}$ can be relatively large, performing an update of all the node features in $\mathcal{G}$ at every training iteration would incur high computational costs. Besides, most of these computations would be of no use as only the updated representations of nodes $\{\mathcal{A}_{i}^{-} \cup a_{i}^{+}\}_{i=1}^{|\mathcal{B}|}$ from batch $\mathcal{B}$ are needed to update the model parameters. Therefore, we build a sub-graph $\mathcal{G}^{sub}$ at each training step that only contains the article nodes from batch $\mathcal{B}$ as well as their $L$-hop neighbors (where $L$ is decided by the number of GNN layers). We then pass that sub-graph to the graph network and use the resulting article representations to compute the loss in \Cref{eq:loss}. Comparably to the node features, the query embeddings are pre-computed offline before training by the query encoder $E_Q^{\boldsymbol{\theta}}(\cdot)$ of our trained bi-encoder.

\section{Experimental Setup \label{sec:experimental_setup}}
In this section, we present the basic setup for experiments. In particular, \Cref{subsec:dataset} describes the dataset we conduct our experiments on, \Cref{subsec:implementation} details our model implementation, \Cref{subsec:baselines} reviews the different baselines we use for comparison, and \Cref{subsec:evaluation} reports the evaluation metrics.

\subsection{Dataset \label{subsec:dataset}}
We conduct experiments on the publicly available Belgian Statutory Article Retrieval Dataset \citep[BSARD]{louis2022statutory}.\footnote{\scriptsize\url{https://huggingface.co/datasets/antoiloui/bsard}} To the best of our knowledge, BSARD is the only SAR dataset that provides the lists of consecutive division headings each article belongs to, which is crucial for building the graph of the legislative structure. The dataset consists of 1,100+ \textsl{French} native questions on various legal topics, as shown in \Cref{tab:bsard_stats}, labeled by skilled experts with references to relevant statutory articles from the Belgian legislation. The retrieval corpus comprises 22,600+ articles collected from 32 Belgian codes covering numerous legal domains. The questions are relatively short and might have several relevant legal articles. We refer readers to the original paper for further data collection and analysis details.

\begin{table}[t]
\centering
\begin{tabular}{lrrr}
\toprule
\textbf{Topic}  & \textbf{Train} & \textbf{Dev} & \textbf{Test} \\ 
\midrule
\family\hspace{0.2cm} Family         & 216            & 56           & 67            \\
\house\hspace{0.2cm} Housing         & 203            & 38           & 66            \\
\money\hspace{0.3cm} Money           & 103            & 35           & 36            \\
\scale\hspace{0.2cm} Justice         & 96             & 25           & 30            \\
\plane\hspace{0.2cm} Foreigners      & 41             & 9            & 13            \\
\old\hspace{0.2cm}   Social security & 27             & 8            & 6             \\
\printer\hspace{0.2cm} Work          & 23             & 6            & 4             \\ 
\midrule
Total           & 709            & 177          & 222           \\ 
\bottomrule
\end{tabular}
\caption{Topic distribution of questions in BSARD.}
\label{tab:bsard_stats}
\end{table}

\subsection{Implementation Details\label{subsec:implementation}}
\paragraph{Model.}
We use the publicly released CamemBERT \citep{martin2019camembert} checkpoint to initialize DSR's query encoder. Due to the specificity of the legal language the article encoder has to deal with, we follow prior work on domain adaptation \citep{gururangan2020stop, rasmus2021multi} and continue pre-training CamemBERT on BSARD statutory articles for 50k gradient steps to adapt it to the target legal domain. We use the resulting domain-specific checkpoint to warm-start the article's first-level encoder. The second-level encoder is a two-layer Transformer encoder of 14M parameters with a similar configuration (i.e., 768-hidden, 3072-intermediate, 12-heads, 0.1 dropout, GeLU). We use max-pooling to aggregate the final chunk representations and cosine as the decomposable similarity function.

\paragraph{Data augmentation.}
Due to the recent success in using synthetic query generation to improve dense retrieval performance \citep{liang2020embedding,ma2021zero,thakur2021heterogenous}, we propose to augment BSARD with synthetic domain-targeted queries. We use a mT5 model \citep{raffel2020exploring} fine-tuned on general domain data from mMARCO \citep{bonifacio2021mmarco} to synthesize queries for our target statutory articles.\footnote{\label{foot:doc2query}\scriptsize\url{https://huggingface.co/doc2query/msmarco-french-mt5-base-v1}} We generate five queries per article, which results in a total of around 118k synthetic queries. We combine the latter with the gold BSARD train samples and obtain an augmented training set of around 122.5k question-article pairs.

\paragraph{Optimization.}
We train DSR for 15 epochs with a batch size of 24 using AdamW \citep{loshchilov2017decoupled} with $\beta_1=0.9$, $\beta_2=0.999$, $\epsilon=$1e-7, weight decay of 0.01, and learning rate warm up along the first 5\% of the training steps to a maximum value of 2e-5, after which linear decay is applied. We then optimize LGE parameters for 10 epochs with a batch size of 512 using AdamW with $\beta_1=0.9$, $\beta_2=0.999$, $\epsilon=$1e-7, weight decay of 0.1, and a constant learning rate of 2e-4. We use 16-bit automatic mixed precision to accelerate training and save memory. Details on our hyperparameter tuning process are given in \Cref{app:gdsr_tuning}.

\paragraph{Hardware \& schedule.}
Training is performed on a single 32 GB NVIDIA V100 GPU hosted on a server with a dual 20-core Intel Xeon E5-2698 v4 CPU \at 2.20GHz and 512 GB of RAM. It takes around 1 day to train DSR and 35 minutes for LGE.

\paragraph{Libraries.}
We implement, train, and tune our models using Transformers \citep{wolf2020transformers}, PyTorch \citep{paszke2019pytorch}, PyTorch-Geometric \citep{fey2019fast}, PyTorch-Lightning \citep{falcon2019pytorch}, W\&B Sweeps \citep{biewald2020wandb}, and DeepSpeed \citep{rasley2020deepspeed}.

\subsection{Baselines \label{subsec:baselines}}
We compare our approach against three strong retrieval systems. As a sparse baseline model, we follow prior work and consider BM25 \citep{robertson1994okapi},\footnote{We use $k_1=2.5$ and $b=0.2$. Details on BM25 hyperparameters tuning are given in \Cref{app:bm25_tuning}.} a popular bag-of-words retrieval function based on exact term matching. We then examine the document expansion technique docT5query \citep{nogueira2019doc2query}, which augments each article with a pre-defined number of synthetic queries generated by a finetuned mT5 model,\cref{foot:doc2query} and then uses a traditional BM25 lexical index from the augmented articles for retrieval. Last, we include the results of a supervised dense passage retrievers \citep[DPR]{karpukhin2020dense} pre-finetuned on more than 90.5k question-context pairs from a combination of three French QA datasets.\footnote{\label{foot:dpr-fr}\scriptsize\url{https://huggingface.co/etalab-ia/dpr-question_encoder-fr_qa-camembert}}

\subsection{Evaluation \label{subsec:evaluation}}
We evaluate model performance using three commonly used ranking measures \citep{manning2008introduction}, namely the macro-averaged recall at different cutoffs (R\at$k$), mean average precision (mAP), and mean r-precision (mRP). Those metrics are further defined in \Cref{app:evaluation_metrics}. We deliberately omit to report the precision\at$k$ given that questions in BSARD have a variable number of relevant articles, which implies that questions with $r$ relevant articles would always have P\at$k < 1$ if $k > r$. Similarly, the mean reciprocal rank (mRR) is not appropriate for BSARD as only the first relevant article would be considered. As some questions might have up to 100 relevant articles, we use $k \in \{100, 200, 500\}$ for the recall\at$k$.

\begin{table*}[t]
\centering
\resizebox{\textwidth}{!}{%
\begin{tabular}{llc|ccc|cc}
\toprule
& \textbf{Model} & \textbf{\#Params} & \textbf{R\at100} $(\uparrow)$ & \textbf{R\at200} $(\uparrow)$ & \textbf{R\at500} $(\uparrow)$ & \textbf{mAP} $(\uparrow)$ & \textbf{mRP} $(\uparrow)$ \\
\midrule
\multicolumn{2}{l}{\textbf{Baselines}} & & & & & & \\
\color{shade}{1} & BM25                                         & -     & 49.3          & 57.3          & 63.0          & 16.8          & 13.6 \\
\color{shade}{2} & docT5query                                   & -     & 51.7          & 59.4          & 65.8          & 18.7          & 15.0 \\
\color{shade}{4} & DPR                                          & 220M  & 77.9          & 81.3          & 88.2          & 45.4          & 39.1 \\
\midrule\midrule
\multicolumn{2}{l}{\textbf{Ours}}  & & & & & & \\
\color{shade}{5} & DSR                                           & 234M & 77.1          & 81.8          & 86.7          & 35.6          & 28.8 \\
\color{shade}{6} & DSR w. domain-adaptive pre-training           & 234M & 79.8          & 83.9          & 88.9          & 39.5          & 31.3  \\
\color{shade}{7} & DSR w. data augmentation                      & 234M & 82.7          & 88.7          & 92.8          & 35.3          & 27.5 \\
\arrayrulecolor{black!30}\midrule
\color{shade}{8} & G-DSR                                         & 262M & \textbf{84.3} & \textbf{90.4} & \textbf{93.1} & \textbf{47.1} & \textbf{40.2} \\
\arrayrulecolor{black}\bottomrule
\end{tabular}%
}
\caption{Retrieval performance on BSARD Test set. The best results are marked in bold.}
\label{tab:results}
\end{table*}

\section{Experiments \label{sec:experiments}}
In this section, we empirically evaluate the effectiveness of our proposed approach against competitive baselines and discuss the main results in \Cref{subsec:results}. Next, we provide an ablation study in \Cref{subsec:ablations} to understand how different design and training options affect our model's performance.

\subsection{Main Results \label{subsec:results}}
\Cref{tab:results} shows retrieval performance on BSARD test set. Although we report model performance on two rank-aware metrics (i.e., mAP and mRP), we emphasize that our approach is specifically aimed at improving the pre-fetching component of a retriever \citep{zhang2021learning} and therefore focuses on optimizing rank-unaware metrics (i.e., R\at$k$).

First, we compare the performance of our proposed G-DSR model$^{\color{shade}\text{(8)}}$ against other well-known retrieval approaches and find it significantly outperforms all of them on SAR. In particular, it improves over the sparse retrieval methods$^{\color{shade}\text{(1,2)}}$ by around 30\% on recall\at$k$ and by more than 25\% on mAP and mRP. It also outperforms a competitive pre-finetuned DPR model$^{\color{shade}\text{(4)}}$ by 6\% on R\at100, 9\% on R\at200, and 5\% on R\at500. However, the latter shows a better performance on rank-aware metrics compared to our DSR models, which we speculate might be due to its extensive pre-finetuning step on three domain-general retrieval datasets, leading the model to a deeper knowledge of the task at hand.

Next, we investigate the influence of different training strategies on the rank-unaware results of our base dense retriever.$^{\color{shade}\text{(5)}}$ We find that DSR's performance is improved when adapting the article text encoder to the legal domain before finetuning on the target data.$^{\color{shade}\text{(6)}}$ Besides, training DSR on a larger dataset containing synthetic domain-targeted queries improves its performance even more.$^{\color{shade}\text{(7)}}$

Finally, our results show that using a GNN model on top of DSR allows to enrich the article representations and leads to the best overall performance.$^{\color{shade}\text{(8)}}$ Interestingly, G-DSR also significantly improves the rank-aware performance of our best performing DSR model by ${\sim}12\%$, suggesting that a GNN could act as an effective re-ranker for SAR.

\subsection{Ablation Study \label{subsec:ablations}}
To further understand how different design choices and training strategies affect the results, we conduct several additional experiments and discuss our findings below.

\paragraph{Alternative pre-trained LMs.}
In addition to CamemBERT, we experiment with several other French or multilingual pre-trained language models to initialize the first-level text encoders in DSR -- namely, mBERT \citep{devlin2019bert}, XLM-R \citep{conneau2020unsupervised}, and ELECTRA-fr \citep{clark2020electra}.\footnote{\scriptsize\url{https://huggingface.co/dbmdz/electra-base-french-europeana-cased-discriminator}} We fine-tune the different warm-started models on BSARD training set and report dev results in \Cref{tab:lm_comparison}. We find that a CamemBERT-initialized DSR model performs best.

\paragraph{Alternative GNNs.}
Additionally to GATv2, we explore different GNN architectures to perform the node feature update -- namely, GCN \citep{kipf2017semi}, GraphSAGE \citep{hamilton2017inductive}, GAT \citep{velickovic2018graph}, and k-GNN \citep{morris2019weisfeiler} -- and summarize the results in \Cref{tab:gnn_comparison}. Our experiments show that using an alternative GNN model does not affect performance much, which suggests that the act of fusing information from neighboring nodes is more important than the way the aggregation is performed.

\paragraph{Similarity and loss functions.}
Besides cosine for scoring pairs of query-article representations, we also experiment with dot-product and Euclidean distance and find both inferior to cosine. As an alternative to negative log-likelihood, we test the triplet loss \citep{burges2005learning} and observe that the latter significantly decreases model performance. More details can be found in \Cref{app:ablation_details}.

\begin{table}[t]
\centering
\resizebox{\columnwidth}{!}{%
\begin{tabular}{lr|cc}
\toprule
\textbf{Pre-trained LM} & \textbf{\#Params} & \textbf{R\at100} $(\uparrow)$ & \textbf{R\at500} $(\uparrow)$\\
\midrule
XLM-R             & 278M  & 59.8            & 78.7 \\
mBERT             & 177M  & 69.2            & 86.6  \\
ELECTRA-fr        & 110M  & 57.6            & 73.7  \\
CamemBERT         & 110M  & \textbf{75.4}   & \textbf{88.5}  \\
\bottomrule
\end{tabular}%
}
\caption{BSARD Dev results of DSR warm-started with different pre-trained word embedding models.}
\label{tab:lm_comparison}
\end{table}

\begin{table}[t]
\centering
\resizebox{\columnwidth}{!}{%
\begin{tabular}{lr|cc}
\toprule
\textbf{Model} & \textbf{\#Params} & \textbf{R\at100} $(\uparrow)$ & \textbf{R\at500} $(\uparrow)$\\
\midrule
GCN         & 14M & 84.4            & 92.0 \\
GAT         & 14M & 84.4            & 92.1 \\
GraphSAGE   & 28M & 84.5            & 92.9 \\
k-GNN       & 28M & 83.8            & \textbf{93.0} \\
GATv2       & 28M & \textbf{84.8}   & 92.3 \\
\bottomrule
\end{tabular}%
}
\caption{BSARD Dev results of LGE with different (3-layer) GNN architectures.}
\label{tab:gnn_comparison}
\end{table}


\section{Related Work \label{sec:related-work}}
Our work operates at the intersection of several research areas, including long document modeling, dense information retrieval, graph neural networks, and legal NLP.

\paragraph{Long document modeling.}
The emergence of deep neural networks for language processing brought new challenges to text encoding, one of which is learning high-quality representations of long documents. For example, \citet{tang2015document} employ a bottom-up approach using CNN and BiLSTM-based hierarchical networks, where sentences are first encoded into vectors, which are then combined to form a single document vector. Similarly, \citet{yang2016hierarchical} build a document vector by aggregating important words into sentence vectors and then aggregating important sentence vectors to document vectors using attention mechanisms. More recently, hierarchical variants of Transformer-based models have been explored for various language tasks, including document classification \citep{mulyar2019phenotyping,pappagari2019hierarchical,chalkidis2019neural,wu2021hierarchical}, summarization \citep{zhang2019hibert}, semantic matching \citep{yang2020beyond}, and question answering \citep{liu2022learning}. In addition to hierarchical attention Transformer-based (HAT) models, several sparse attention Transformers (SAT) have been introduced to reduce the computational complexity of the model, thus allowing to process sequences longer than 512 tokens \citep{child2019generating,beltagy2020longformer,zaheer2020bigbird}. However, \citet{chalkidis2022exploration} show that a pre-trained HAT model performs comparably or better than an equally-sized SAT model across several downstream tasks while being substantially faster and less memory intensive. Recently, other non-Transformer-based approaches have been proposed for efficient long sequence processing based on structured state spaces \citep{gu2022efficiently,gupta2022diagonal}.

\paragraph{Dense information retrieval.}
Traditionally, lexical approaches such as TF-IDF and BM25 \citep{robertson1994okapi} have been the \textsl{de facto} standard for textual information retrieval due to their robustness and efficiency. However, these approaches suffer from the lexical gap problem \citep{berger2000bridging} and can only retrieve documents containing keywords present in the query. To overcome this limitation, recent work relies on neural-based architectures to capture semantic relationships between pairs of texts \citep{lee2019latent, karpukhin2020dense,yang2020multilingual,xiong2021approximate}. These models map queries and documents into dense vector representations and calculate a relevance score by the similarity of the vectors \citep{gillick2018end}, which allows the document representations to be pre-computed and indexed offline for inference. The dense retrieval approach was recently extended by hybrid lexical-dense methods, which aim to combine the strengths of both approaches \citep{joon2019real,gao2021complement,luan2021sparse}. We refer the readers to \citet{yates2021pretrained} for a survey on neural information retrieval.

\paragraph{Graph neural networks.}
Graph neural networks (GNNs) capture the topological relationships among the nodes of a graph using an information diffusion mechanism that propagates node features according to the underlying graph-structured data \citep{scarselli2009graph}. These models have shown their effectiveness and flexibility in a wide variety of NLP tasks, including text classification \citep{lin2021bertgcn,yu2022jaket}, relation extraction \citep{zhang2018graph,li2020graph,carbonell2020named}, and question answering \citep{cao2019question,xu2021fusing}. Recently, GNNs have been employed for document retrieval to enhance the vector representations by leveraging the topological structure of the documents, where nodes are passages from a document and edges are relations between these passages \citep{xu2021contrastive,zhang2021graph,albarede2022passage}.

\paragraph{Application to the legal domain.}
In recent years, the legal domain has attracted much interest in the NLP community, both for its challenging characteristics and massive volumes of textual data \citep{chalikidis2019deep,zhong2020how}. Researchers see it as an opportunity to develop novel automated methodologies that can reduce heavy and redundant tasks for legal professionals while providing a reliable, affordable form of legal support for laypeople \citep{bommasani2021opportunities}. Earlier techniques for legal information retrieval were mainly based on term-matching approaches \citep{kim2017two,tran2018jnlp}. Recently, a growing number of works have used neural networks to enhance retrieval performance, including word embedding models \citep{landthaler2016extending}, doc2vec models \citep{sugathadasa2018legal}, CNN-based models \citep{tran2019building}, and BERT-based models \citep{nguyen2021jnlp,chalkidis2021regulatory,althammer2022parm}. To the best of our knowledge, we are the first to exploit the structure of statute law with GNNs to improve the performance of dense retrieval models.

\section{Conclusion \label{sec:conclusion}}
In this paper, we introduce G-DSR, a novel approach for statutory article retrieval (SAR) that leverages the topological structure of legislation to improve retrieval performance. Specifically, G-DSR enriches the article representations of a dense retriever designed for long document retrieval by employing a graph neural network that uses the organization of statute law to learn knowledge-rich cross-article embeddings. Experiments show that G-DSR outperforms competitive baselines on a real-world expert-annotated SAR dataset. We also include a detailed analysis to motivate our design choices and training strategies.

\section*{Limitations \label{sec:limitations}}
While our approach  performs well on statutory article retrieval, it comes with several limitations that provide avenues for future work.

First, experimental results are based on questions and labels drafted by legal professionals. It is possible that other legal professionals would draft the questions differently or, less likely yet possible, that they would deem different statutory provisions relevant. This raises the question of to what extent similar results would be obtained if the model were trained on a different dataset, for instance, based on other experts or domains, hence testing the approach's generalizability. The main challenge in this regard is obtaining data, as organizations are unlikely to share or even collect similar data.

Second, our proposed methodology was evaluated exclusively on the Belgian legislation, whose laws are organized in a hierarchical manner where the deeper the divisions, the more closely related the legal concepts of the articles under them. Although we believe our approach could be applied to most, if not all, jurisdictions that rely on statute law (including both civil and common law countries), different jurisdictions may have different organizations of their legal provisions, which could potentially affect the model's performance. It is also worth mentioning that the dataset used for evaluation comes with a linguistic bias as Belgium is a multilingual country with French, Dutch, and German speakers, but the provided provisions are only available in French. Studying the applicability and impact of the present work to other jurisdictions and languages is an exciting research direction that is challenging in practice due to the scarcity of high-quality multilingual statute retrieval datasets.

Then, our approach currently considers the topological structure of legislation for modeling the inter-article dependencies, which implies that information is aggregated between direct neighboring articles only while those from more distant sections are completely ignored. Nevertheless, it is common for articles to cite other articles from different sections or even different statutes. Therefore, we believe that considering richer legal graph structures, especially legal citation networks, could increase effectiveness even more. However, building such citation networks from raw texts requires a considerable text-processing effort.

Finally, although G-DSR shows promise for statutory article retrieval, it is not yet ready for practical use in the real world. One issue is that our model is designed to be an effective pre-fetcher, optimizing recall such that all articles relevant to a question appear in an unordered filter set of size $k$ ($k$ being relatively large). However, in practice, users would expect a high-quality retrieval system to not only find these relevant articles but also to sort them by decreasing order of importance, requiring an adequate re-ranker. Then, it is essential to recognize that while access to relevant legal provisions is a necessary step in helping the general public solve their legal issues, it is not a sufficient condition on its own as laypeople may still struggle to understand the legal jargon and apply the provisions to their specific situations. Ideally, the tool to be made accessible to the public should consist of a two-stage framework: (i) a legal provision retriever, which selects a small subset of relevant legal articles in response to a given question, and (ii) a legal-to-natural translator or summarizer, which examines the retrieved articles and generates an answer in natural language. In the present work, we chose to focus on the first stage of this framework and leave the second for future work.

\section*{Ethics Statement}
The scope of this work is to provide a new methodology along with extensive experiments to drive research forward in statutory article retrieval. We believe the latter is an important application field where more research should be conducted to improve legal aid services and access to justice for all. We do not foresee situations where the use of our methodology would lead to harm \citep{tsarapatsanis2021ethical}. Nevertheless, although our goal is to improve the understanding of the law by those who suffer from legal information asymmetry, it cannot be excluded that the technology presented here could exacerbate inequality if states, companies, or lawyers benefit more from its use than the intended beneficiaries (i.e., citizens, consumers, or employees).

\section*{Acknowledgments}
This research is partially supported by the Sector Plan Digital Legal Studies of the Dutch Ministry of Education, Culture, and Science. In addition, this research was made possible, in part, using the Data Science Research Infrastructure (DSRI) hosted at Maastricht University.

\bibliographystyle{packages/acl_natbib}
\bibliography{refs.bib}

\appendix
\section*{Appendix}

\section{Preliminary Studies \label{app:preliminary_studies}}

\subsection{How important is the structure of law for statutory article retrieval? \label{app:experts_approach}}
To better understand the reasoning skilled humans follow when retrieving statutory articles, we ask several legal experts to retrieve relevant articles to questions sampled from the Belgian Statutory Article Retrieval Dataset \citep[BSARD]{louis2022statutory}. We deliberately choose experts unfamiliar with Belgian law and thus have no past knowledge of the location of articles covering a particular subject. In what follows, we summarize the approach these experts use.

First, they determine whether the issue involves either \textsl{public} or \textsl{private} law. To distinguish between the two, it is necessary to identify to whom the rules apply (i.e., the parties involved in the issue that hold the rights or duties). Generally speaking, public law deals with issues that affect the general public or state (i.e., society as a whole), whereas private law deals with issues that affect individuals, families, and businesses. This first step allows the experts to make an initial selection among the codes of law, which generally relate to only one of either public or private law. 

Next, the experts refine their search by determining the field of law (e.g., contract law), followed by the sub-field (e.g., tenant law), and so on, until a set of potentially relevant codes is created in the question's domains. The experts then focus on the table of contents of one of the selected codes and undertake a hierarchical search that starts from the book's headings and progressively extends to its titles, chapters, and sections. This step makes it possible to filter out many irrelevant articles by analyzing the connection between the question's subject and the different sections' headings. 

Finally, the experts explore the articles within the sections deemed potentially relevant to the question in search of the expected answer. If the experts realize that the chosen direction is a dead end, they return to the previous higher level of the structure, choose another potentially relevant direction, and narrow their search from there. 

From this study, we conclude that legal experts rely heavily on the structure of law when retrieving relevant articles to a legal question, which indicates that the different divisions' headings carry valuable information that retrieval systems should consider.

\subsection{How related are neighboring articles in statute law? \label{app:neighboring_articles}}
In statute law, the sense of a given article is not necessarily self-contained by itself but instead spans across different articles from the same or even different sections. To confirm this, we study to what extent consecutive articles (as they appear in the statute books) address similar subjects.

\begin{figure}[t]
    \centering
    \includegraphics[width=1\linewidth]{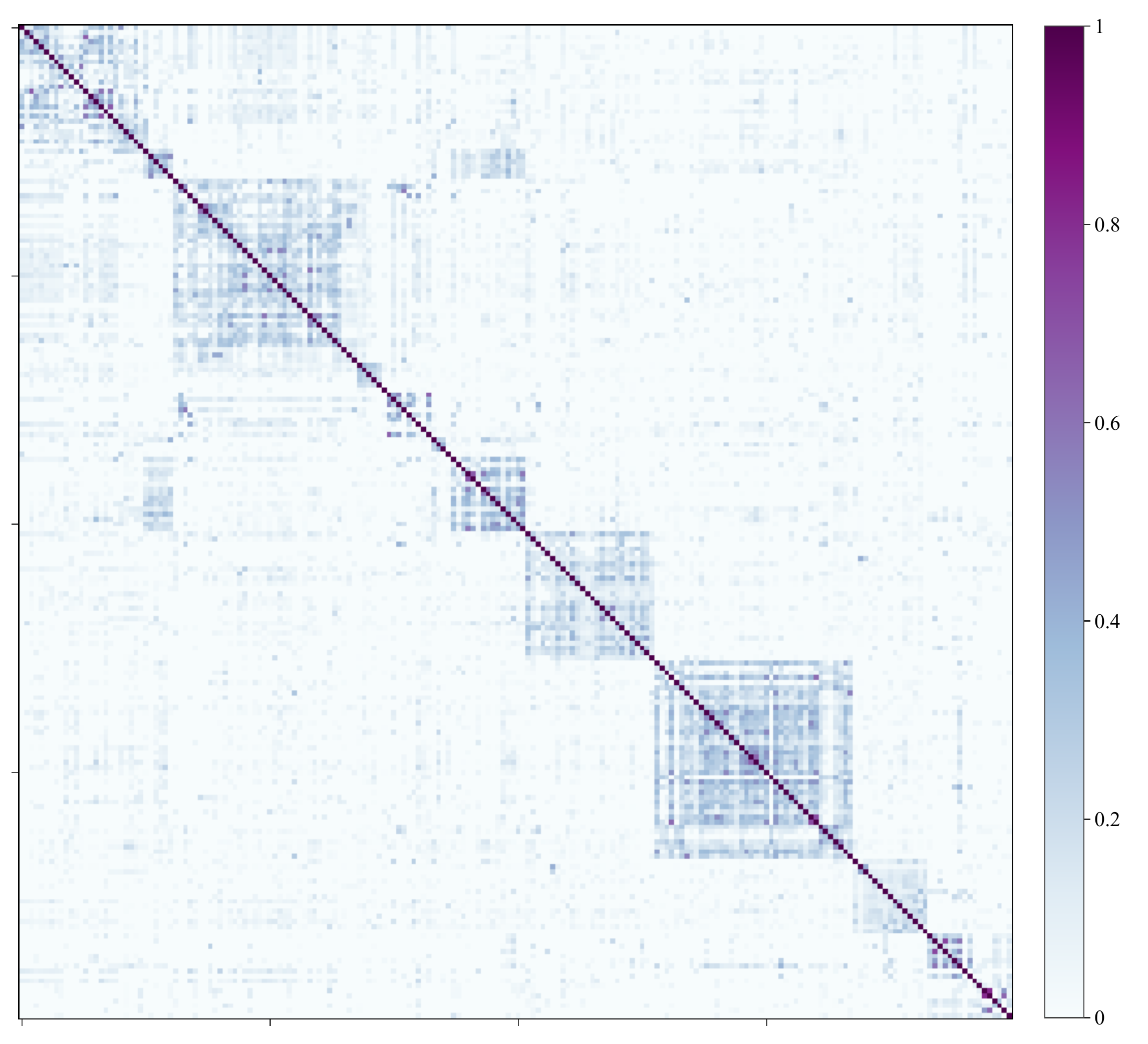}
    \caption{Cosine similarities between TF-IDF representations of 200 consecutive articles from the Belgian Civil Code, taken from the Belgian Statutory Article Retrieval Dataset \citep[BSARD]{louis2022statutory}.}
    \label{fig:article_correlation}
\end{figure}

We consider the Belgian Civil Code, which is the book whose articles are most cited in BSARD, and randomly sample sets of 200 consecutive articles out of it. We then normalize the articles by lowercasing, lemmatizing, and removing stop-words, punctuation, and numbers. Finally, we compute the cosine similarities between the TF-IDF representations of all articles from a given set. \Cref{fig:article_correlation} shows a heatmap of article similarities for such a set. We see that consecutive articles do indeed cover similar topics, suggesting that the information in a given article is likely to be complementary to that in its neighboring articles. Therefore, we assume that neighboring articles should be considered to capture an article's whole meaning.

\begin{table}[ht]
\centering
\resizebox{\columnwidth}{!}{%
\begin{tabular}{llrr}
\toprule
            & \textbf{Hyperparam}              & \textbf{DSR} & \textbf{LGE} \\
\midrule
\multicolumn{2}{l}{\textbf{Model}}    &              &              \\
            & Maximum Chunk Length             & 128          & -            \\
            & Maximum Document Length          & 1024         & -            \\
            & Pooling Strategy                 & max          & -            \\
            & \#Layers                         & -            & 3            \\
\midrule
\multicolumn{2}{l}{\textbf{Loss}}     &              &              \\
            & Temperature                      & 0.01         & 0.01         \\
            & Similarity                       & cos          & cos          \\
\midrule
\multicolumn{2}{l}{\textbf{Training}} &              &              \\
            & Batch Size                       & 24           & 512          \\
            & Weight Decay                     & 0.01         & 0.1          \\
            & Max Epochs                       & 15           & 10           \\
            & Peak Learning Rate               & 2e-5         & 2e-4         \\
            & Learning Rate Decay              & Linear       & Constant     \\
            & Warmup ratio                     & 0.05         & 0.0          \\
            & AdamW $\epsilon$                 & 1e-7         & 1e-7         \\
            & AdamW $\beta_1$                  & 0.9          & 0.9          \\
            & AdamW $\beta_2$                  & 0.999        & 0.999        \\
            & Gradient Clipping                & 1.0          & 1.0          \\
\bottomrule
\end{tabular}%
}
\caption{Training hyperparameters for DSR and LGE.}
\label{tab:hyperparameters}
\end{table}

\section{G-DSR Hyperparameter Tuning \label{app:gdsr_tuning}}
We conduct hyperparameter tuning using Bayes search based on performance on BSARD development set, measured with the macro-averaged R\at200. Due to limited computational resources, we train our models on BSARD training set only -- which takes approximately 1 hour and 15 minutes for DSR and around 5 minutes for LGE -- and use the constrained search spaces described below.

\paragraph{DSR grid search space:}
\begin{itemize}[noitemsep]
    \item batch size: \{8, 16, 24, 32\}
    \item learning rate: \{5e-5, 4e-5, 3e-5, 2e-5, 1e-5\}
    \item weight decay: \{0, 0.1, 0.01, 0.001\}
    \item max chunk length: \{64, 128, 256, 512\}
    \item max document length: \{1024, 2048\}
    \item pooling strategy: \{mean, max\}
    \item similarity function: \{dot-product, cosine, L2\}
    \item temperature: \{0, 0.1, 0.01, 0.001\}
\end{itemize}

\paragraph{LGE grid search space:}
\begin{itemize}[noitemsep]
    \item batch size: \{8, 16, 32, 64, 128, 256, 512\}
    \item learning rate: \{2e-2, 2e-3, 2e-4, 2e-5, 2e-6\}
    \item weight decay: \{0, 0.1, 0.01, 0.001\}
    \item \#layers: \{1, 2, 3, 4\}
\end{itemize}

In total, we run 100 hyperparameter search trials for both DSR and LGE. The optimal hyperparameters, shown in \Cref{tab:hyperparameters}, are used to re-train the models combining both train and development sets for a final evaluation on the test set.

\section{BM25 Hyperparameter Tuning \label{app:bm25_tuning}}
\begin{figure}[t]
    \centering
    \includegraphics[width=1\linewidth]{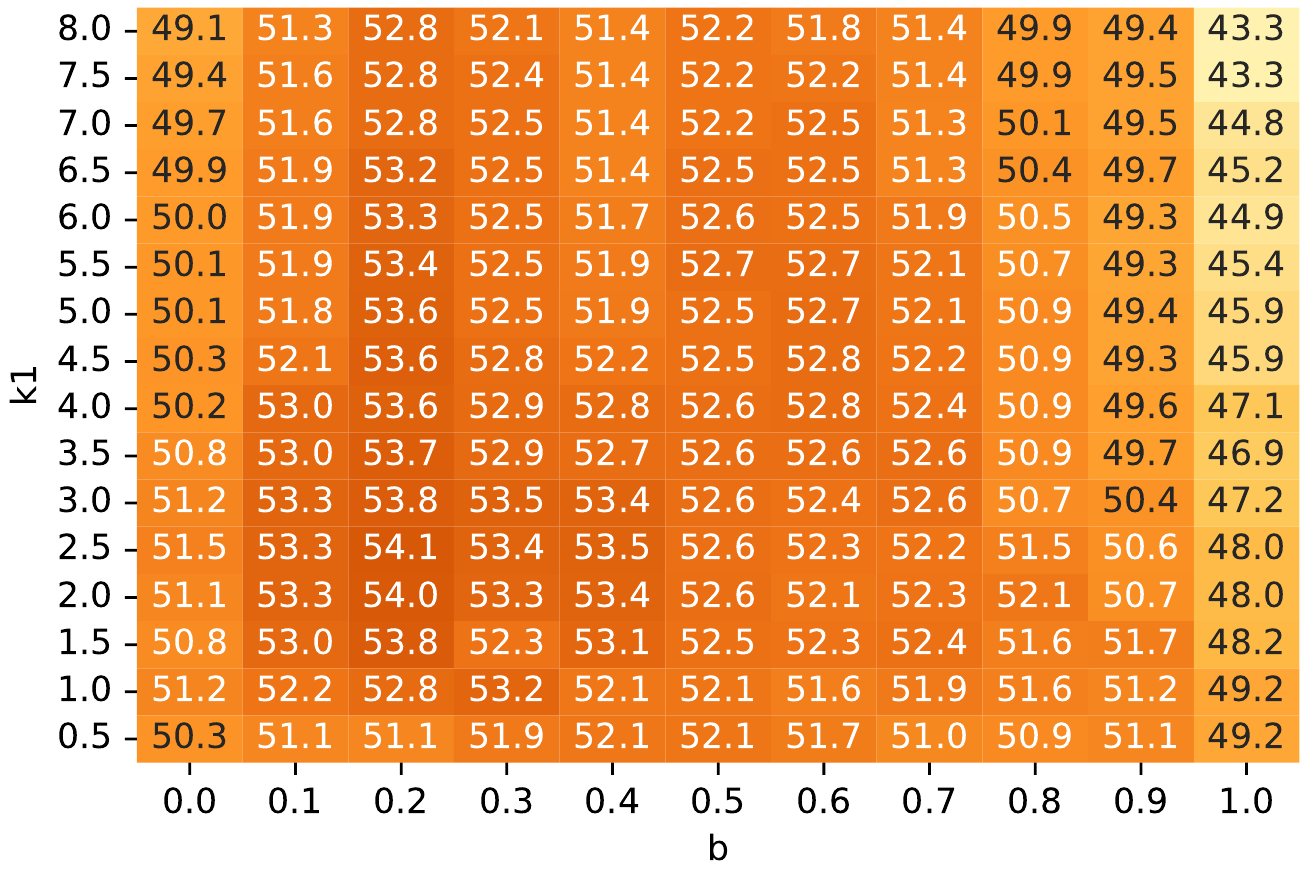}
    \caption{Heatmap showing BM25 results on BSARD Dev set for different values of $k_1$ and $b$.}
    \label{fig:bm25_tuning}
\end{figure}

Following \citet{chalkidis2021regulatory}, who show that BM25 performance is highly dependent on adequately choosing the $(k_1, b)$ values for the task at hand, we perform a hyperparameter grid search on BSARD development set and plot the results in \Cref{fig:bm25_tuning}. We observe that, in the case of SAR, the best performance is obtained with $k_1=2.5$ and $b=0.2$. Therefore, we use these values for the final evaluation on BSARD test set.

\section{Evaluation Metrics \label{app:evaluation_metrics}}
Let $\operatorname{rel}_{q}(a) \in \{0,1\}$ be the binary relevance label of article $a$ for question $q$, and $\langle i, a\rangle \in \mathcal{F}_q$ a result tuple (article $a$ at rank $i$) from the filter set $\mathcal{F}_q \subset \mathcal{C}$ of ranked articles retrieved for question $q$.

\paragraph{Recall.}
The \textit{recall} is the fraction of relevant articles retrieved for query $q$ w.r.t. the total number of relevant articles in the corpus $\mathcal{C}$, i.e.,
\begin{equation}
    \operatorname{R}_{q}=\frac{\sum_{\langle i, a\rangle \in \mathcal{F}_q} \operatorname{rel}_{q}(a)}{\sum_{a \in \mathcal{C}} \operatorname{rel}_{q}(a)}.
\end{equation}
When computed for a filter set of size $k=\left|\mathcal{F}_q\right|\ll|\mathcal{C}|$, i.e., at a certain cutoff and not on the entire list of articles in $\mathcal{C}$, we report the metrics with the suffix ``\at\textit{k}''.

\paragraph{R-Precision}
The R-Precision is the proportion of the top-$R$ retrieved articles that are relevant to query $q$, where $R$ is the total number of relevant articles for $q$, i.e., 
\begin{equation}
    \operatorname{RP}_{q}=\frac{\sum_{\langle i, a\rangle \in \{\mathcal{F}_q\}_{i=1}^{R}} \operatorname{rel}_{q}(a)}{R}.
\end{equation}

\paragraph{Average Precision.}
The \textit{average precision} is the mean of the precision value obtained after each relevant article is retrieved, that is
\begin{equation}
    \operatorname{AP}_{q}=\frac{\sum_{\langle i, a\rangle \in \mathcal{F}_q} \operatorname{P}_{q, i} \times \operatorname{rel}_{q}(a)}{\sum_{a \in \mathcal{C}} \operatorname{rel}_{q}(a)},
\end{equation}
where $\operatorname{P}_{q, j}$ is the \textit{precision} computed at rank $j$ for query $q$, i.e., the fraction of relevant articles retrieved for query $q$ w.r.t. the total number of articles in the retrieved set $\{\mathcal{F}_q\}_{i=1}^{j}$:
\begin{equation}
    \operatorname{P}_{q, j}=\frac{\sum_{\langle i, a\rangle \in \{\mathcal{F}_q\}_{i=1}^{j}} \operatorname{rel}_{q}(a)}{\left|\{\mathcal{F}_q\}_{i=1}^{j}\right|}.
\end{equation}
We report the macro-averaged \textsl{recall} at various cutoffs (R\at$k$), \textsl{mean Average Precision} (mAP), and \textsl{mean R-Precision} (mRP), which are the average values over a set of $n$ queries.

\begin{table}[t]
\centering
\resizebox{\columnwidth}{!}{%
\begin{tabular}{ll|ccc}
\toprule
\textbf{Loss} & \textbf{Similarity} & \textbf{R\at100} $(\uparrow)$ & \textbf{R\at200} $(\uparrow)$ & \textbf{R\at500} $(\uparrow)$ \\
\midrule
\multirow{3}{*}{Cross-entropy} & Cosine    & \textbf{75.0}  & \textbf{80.7} & \textbf{86.7} \\
                               & Dot       & 21.6           & 37.1          & 56.0 \\
                               & Euclidean & 43.9           & 58.1          & 71.6 \\
\midrule
\multirow{3}{*}{Triplet}       & Cosine    &  5.4           & 9.9           & 17.2 \\
                               & Dot       &  4.3           & 6.4           & 10.3 \\
                               & Euclidean &  9.0           & 14.3          & 25.2 \\
\bottomrule
\end{tabular}%
}
\caption{BSARD Dev results of DSR trained using different similarity and loss functions.}
\label{tab:sim_and_loss}
\end{table}

\section{Ablation Details \label{app:ablation_details}}
Besides cosine similarity and negative log-likelihood (NLL) loss, we also test the dot-product and Euclidean (inverse of distance is taken as similarity measure) as well as the triplet loss. The temperature for the NLL loss is set to 0.01, and the margin value of the triplet loss is set to 1. We report the results on BSARD development set in \Cref{tab:sim_and_loss}. For a fair comparison, all models are trained for 15 epochs with a batch size of 24, weight decay of 0.01, warm-up proportion of 0.05, an initial learning rate of 2e-5, and a linear decay learning rate schedule.

\end{document}